\documentclass[a4paper,notitlepage, 12pt]{article}
\usepackage{amsfonts}
\usepackage{amssymb}
\usepackage{amsmath}
\usepackage{bm}
\usepackage{graphicx}
\def\one{{\hbox{1\kern-.8mm l}}}

\usepackage{color}
\usepackage{graphics}

\newcommand{\beq}{\begin{equation}}
\newcommand{\eeq}{\end{equation}}
\newcommand{\be}{\begin{eqnarray}}
 \newcommand{\ee}{\end{eqnarray}}
\newcommand{\ov } {\over }

\begin{document}
\begin{titlepage}
\bigskip\begin{flushright}
NORDITA-2009-20\\
\end{flushright}
\begin{center}
\vskip 4cm
\Large{Self-interacting fundamental strings and black holes}
\end{center}
\vskip 1cm
\begin{center}
\large{Diego Chialva \\ 
       {\it  Nordita Institute,
       AlbaNova University Centre,
   Roslagstullsbacken 23
   SE-106 91 Stockholm, Sweden} \\
        \tt{chialva@nordita.org}}

\end{center}
\date{}

\pagestyle{plain}
\vskip 3cm
\begin{abstract}
We study the size distribution of very massive close string states
and the typical string configuration
as one slowly increases the string
coupling,
both in the case of zero and of non-zero Neveu-Schwarz charges. 
The computations are performed rigorously in string theory, starting
from quantities that are well-defined in the theory and therefore
clarify previous works on the subject which were based on various approximation
techniques. 

We find that, starting from a value of the coupling in
agreement with the one predicted by the black hole correspondence
principle, the string ensemble 
is dominated  in any dimensions 
by compact states whose size is within the correspondent black
hole horizon radius, which is of the order of
the string scale at the black hole/string transition/matching point.

\end{abstract}
\end{titlepage}
\newpage

\tableofcontents

\setcounter{section}{0}
\section{Introduction}

Black holes obey a set of laws formally identical to the
thermodynamic ones. In particular to every black hole are
associated an entropy and a temperature.
These have to be accounted for by an underlying microscopical theory, which
should provide us not only with an interpretation of black holes
entropy, but also with the possibility of deriving the whole set of the
thermodynamic laws from first
principles.

According to Bekenstein's
principle, the black hole entropy is proportional to the area of the
horizon (plus corrections\footnote{The more general
formula for the entropy of a black hole, including higher
derivative terms, is Wald's one \cite{Wald}.}); for a quantum
statistical ensemble,  
instead, the entropy
is defined as the logarithm of the number $G$ of microstates:
 \beq
 S = \ln(G).
 \eeq

Relating the two definitions represents the entropy issue. It is
necessary to individuate the correct ensemble of microstates accounting
for $S$: we need therefore a quantum gravity theory. 
String theory represents probably the best candidate nowadays for such
a theory and it has a general principle (known as the
``String-Black holes correspondence principle'' \cite{corresp,
  HorPol96}) individuating those 
microstates. 

A key role in the correspondence principle is
played by the relationship between the Schwarzschild radius of black holes
and the string length scale, the two characteristic scales set by
black hole Physics and string theory. At the matching point these
two should be equal \cite{HorPol96}
and we would expect only string states whose size
is within the Schwarzschild radius to be related to the
black hole description\footnote{Consider for example the gravitational
  binding to mass ratio, or the fact that
the metric sourced by the microstates should differ from the one of a black hole
only at distance lower than the horizon radius.}. We need therefore to
be able to study 
properties of sets of states at given mass, charge, angular
momentum but also ``size''. 

In this work we want to focus on two aspects of this problem: 1) 
on properly computing the
entropy of microstates in
string theory in terms of their mass,
(Neveu-Schwarz) charges and size, 2) on determining how the
configuration of string states ensembles (their distribution in terms
of size) changes by varying (adiabatically) the string coupling.
 
 This has never been done before rigorously in
string theory formalism, because of two principal reasons.
On one hand, it
is not straightforward, in the quantum theory of strings, to define an
operator  
measuring the (average) size of string states and this complicates the
analysis already at tree-level. On the other hand, taking into account
string self-interactions requires one-loop string computations which
are difficult to be performed. Because of all this,
the partition function for single-string states
constrained\footnote{We will use the microcanonical
  ensemble, as explained in the following.} in both 
mass, charges and size is unknown both at
tree-level (free string) and at non-zero string coupling.
The main result of this work will therefore be
to compute it in a well-defined way and to obtain the entropy from it.

Note that the free-string entropy as a function of mass, charge and size
would now receive corrections in both the non-BPS and the BPS cases
(renormalization of the size), however we will concentrate on the
self-energy contribution for the non-BPS case.
In particular, the one-loop corrections to our tree-level results will
be obtained and discussed elaborating on the results in
\cite{chialvamassshift}. 
 
A part from the interest in relation with black holes, 
studying string configuration (size) and entropy will
also clarify the link between the 
string theory and the random-walk pictures in full details\footnote{ In
the past, instead, the only quantities computed within String Theory
and compared with the random-walk picture were just the root mean
square radius and some density-density correlators, as we will mention.}.

We
consider closed string, and perform our 
analysis both in the bosonic and in the superstring theory (type IIB or IIA).
We will deal
with states carrying both zero and non-zero charges of the
Neveu-Schwarz type (Kaluza-Klein momentum and winding numbers).

The approach we follow goes {\em beyond} those used so far, which
assumed some sort of approximation in the description of strings 
(random-walk, polymers, thermal 
scalar, \ldots) and led sometimes to results actually being in
contradiction (see \cite{HorPolSelf, DamVenSelf}).
Indeed, we will
compute the partition function from 
first principles consistently within String Theory.

We begin in Section \ref{principle},
by reviewing the String-Black Holes Correspondence Principle. Special emphasis
is given to the role played by the value of the horizon radius of the
black hole and the corresponding requirements for the size of the
string microstates. This section establishes the basis and motivation
for the present study. We also review the existing literature
on the subject and the controversial and obscure points.

\ref{microensemble} and
\ref{measuresetup} are introductory sections where we
discuss and specify our statistical ensemble of closed string
states and the definition of ``size'' of a string\footnote{An
  important part of the proof that we 
are indeed computing the correct partition function will be presented
later, in
section \ref{numberstatessection}.}. 

The rest of the paper is divided in two parts: the first one (section
\ref{partifuninstr}) dealing with the size distribution of highly
excited {\em free} string states, the second one (section
\ref{massshiftsec}) focusing on the corrections to the size distribution
due to self-interaction of the string.

Finally, we comment and conclude.

\section{The String-Black Holes correspondence principle}\label{principle}

In this section, for simplicity, we consider the case with zero
charges\footnote{For non-zero charges the Schwarzschild radius
  (horizon radius) in the string frame and that in the Einstein frame
  are different and care must be used in stating the correspondence
  principle, see \cite{HorPol96} and footnote \ref{gravdress}.}.
String theory and black holes' physics set two
characteristic length scales:
 \begin{tabbing}
 ~~~~$R_{bh}$ \hspace{1.4cm} \=
 the black hole horizon radius 
 (Schwarzschild radius) \\
 ~~~~$l_s=\sqrt{\alpha'}$ \> the string length scale
 \end{tabbing}
so that
 \beq
 \begin{cases}
 \text{if}\quad R_{bh}\gg l_s & \text{general relativity description is
 reliable} \\
 \\
 \text{if} \quad R_{bh} \lesssim  l_s & \text{strings feel space-time as flat,
 $\alpha'$-corrections are important,}\\
                      & \text{string theory description is reliable}  \\
 \end{cases} \nonumber
 \eeq

The  ``String-Black Holes Correspondence
Principle'' \cite{corresp, HorPol96} says that a black hole is described by 
an ensemble of excited string
and/or D-brane states (depending on the type of charges the black hole
possesses). 

There are two possible
interpretations of the Principle:
 \begin{itemize}
 \item a {\bf physical process} (Hawking radiation) where the black
 hole decreases its mass $M$, therefore reducing the value of its
 Schwarzschild
 radius\footnote{We are in
$d=D-1$ extended spatial dimensions. In perturbative regime, which
   will be the one we will work in, we relate Newton's constant
  to the string length
       as $G_N\sim g_s^2 (\alpha')^{{d-1 \ov 2}}$, considering
       compactified dimensions with curvature radii $r_i \sim \sqrt{\alpha'}$.} 
$R_{bh}\sim (G_N M)^{{1 \ov d-2}}$ until $R_{bh}\sim l_s$
 where a transition 
to an excited string states takes place. 
In this case the (closed) string coupling 
 $g_s $ is fixed, while $M$ varies.
 \item {\bf two complementary descriptions} valid in different regimes
 at equal mass. In this case $g_s $ varies, while $M$ is fixed. This
 is the approach that we will use in this work.
 \end{itemize}

The two descriptions can be compared at determined values of the
coupling when they are both valid.
The possibility of equating the black hole entropy (proportional to a
power of its mass) and the 
string one (proportional to the square root of its mass) relies on the
fact that the first is constant in Plank units, the second in string
ones and therefore the entropies match at a determined value 
of the string coupling.
There, at 
the matching point, entropy and mass\footnote{And charges when
  they are present.} of the black hole
and the string are equal and it is found that, in units of $\alpha'$ 
\cite{HorPol96},
 \beq 
 g_s \sim M^{-{1 \ov 2}}, \qquad R_{bh} \sim l_s
 \eeq
independently of the number of dimensions.
Since we are to consider very
massive string states, this value for the string coupling turns out to
be sufficiently small to allow perturbation theory.

We would expect that only states 
whose size is within the black hole horizon radius can be related to
the black hole at 
the transition/matching point. 
It is therefore interesting and important to find the entropy of
string microstates 
depending on both mass and size.

The main novel result of this work is the definition and the
computation in a fully consistent way within string theory of the
partition function (and therefore entropy)
for states constrained in mass, (charge) and size\footnote{It would also be
  possible to characterize the states 
  through their angular momentum, obtaining the entropy as a function
  of mass, charge, size and angular momentum, following \cite{SussRuss}.} 
(in the microcanonical ensemble) and the study of the string
configuration when the coupling $g_s$ is slowly increased. 

In the past, a few attempts have been made to study
such issues: \cite{KalyanaRama},
\cite{Khuri} and especially \cite{HorPolSelf}, \cite{DamVenSelf}
(see also
\cite{Cornalba}). 

In 
\cite{HorPolSelf}, it was
employed a thermal scalar formalism, interpreting the size of
the bound states of a certain scalar field as the size of the excited
{\em open} string states. The thermal scalar is a formal device
capable to give us some 
statistical information about the string system (string
gas). Nevertheless its relation to the string states remains
open. In 
particular, interpreting the
size of bound states of the thermal scalar as the size of the string
states in Minkowski
space is not obvious since, as the authors of \cite{HorPolSelf} remark
themselves,  
the thermal scalar has no dynamical meaning. However we will show how
the results in \cite{HorPolSelf}, taking into account the differences 
between open and closed strings, compare to ours.

The polymer and string bit picture for strings was the
approximation used in  \cite{KalyanaRama}, \cite{Khuri}. This
is related to the random-walk one in the idea of the string depicted
as formed by an ensemble of ``bits''. But these models have many
open issues, in particular for the superstring (see the discussion in
\cite{Bergman} and in \cite{KalyanaRama} itself). 

The results obtained in \cite{DamVenSelf} merge instead 
physical intuition with computations in a model of (bosonic open)
``strings'', believed 
to be valid in a large number of dimensions ($d \gg 1$), which did
not take into account any Virasoro constraint. 
The results of  \cite{DamVenSelf}, if correct, 
are somehow puzzling and raise some
concern about the viability of perturbation theory in string theory,
as we will now discuss.

The tree-level logarithm of the string partition function for states
at large masses $M=\sqrt{N}$, resulting 
from the computation in their model (formula (2.28) in
\cite{DamVenSelf}, using formulas (2.10, 2.27, 2.28)) reads
 \beq
 \log(Z) = 
  \sqrt{{\pi^2 d \ov 6} N}\left(1-{9 \ov 4 }{1 \ov \bar R_s^2}
  \right)
 \eeq
and is obtained at
large mass
through  an expansion for $\delta \sim {l_s \ov
  R_s^2} \ll 1$, therefore is a result valid for large string
sizes\footnote{The
  result is obtained through an expansion for large $R_s$, as we said,
  and not limited in any
way by the value of the mass of
the states, see formula 2.25 in \cite{DamVenSelf}. A posteriori it is
declared that 
the result is valid only for $R_s^2 < M$ but there is no reason from
the computation for doing so.} $R_s$. 

This result is
puzzling, since for large size and masses we would expect the string
to behave like a random-walk. Indeed, the authors of \cite{DamVenSelf}
need to combine by hand the result from the computation with a corrective
factor $e^{-R_s^2 M^{-2}}$ to the partition function in order to
obtain physically reasonable results. The problem is that the operator
measuring the size of string and the Fock space used in
\cite{DamVenSelf} are not 
well-defined. The same (as for the operator) happens in
\cite{Cornalba}.

For what concerns their estimate of the one-loop corrections
(squared average mass-shift) for open self-gravitating strings, let us consider
their result from formula (3.26) and the line above (3.25). At
small string coupling they find a typical size
$R_{av} \sim \sqrt{M}$ for the average string. Therefore the result in
\cite{DamVenSelf} for the average squared mass
shift is (using the notation there)
 \beq \label{damvenmassshift}
 \delta M^2 = -c g_s^2 M^3 M^{{2-d \ov 2}}=
 -c g_s^2 M^2 M^{{4-d \ov 2}} \,.
 \eeq
We see that for $d=3$ this result means that at a
given\footnote{Recall that $g_s$ in string theory is not a free
  parameter, but depends  
  on the dilaton vacuum expectation value, although in this work we use it
as a useful parameter for investigating the string configuration.} $g_s$,
for states with sufficiently
large $M^2$, the correction to the mass squared will be larger than
the original value (that is $\delta M^2 >  M^2$). 
This would mean that we cannot apply perturbation
theory consistently on the whole string spectrum, in particular not in the
limit of large masses.

It is therefore
fair to say that the results in \cite{DamVenSelf} need to be verified
both at tree and one-loop level\footnote{The authors of
  \cite{DamVenSelf} 
  acknowledge the difficulties and their investigation is on a
  physically motivated ground. Nevertheless it is important to verify
  it.}.

We will perform our calculations rigorously in full-fledged string theory.
Our conventions, here and in the following, are
 \be
 \alpha' = 4,\footnotemark && \quad 
 D=d+1 \,\,\text{large space-time dimensions} \nonumber\\
 g_s/g_o =&& \text{closed/open string coupling}. \nonumber
 \ee
\footnotetext{Our results will be written in string units.}

Furthermore, quantities with a ``$c$'' subscript will refer to closed
strings, whereas those 
with an ``$o$'' subscript will indicate open strings.
Finally,
a tilde or a subscript $R$ refer to the
right-moving sector of the 
close string, a subscript $L$ or no tilde refer to the
left-moving one.

The computation of the entropy of strings
is ultimately connected also with the Hagedorn transition in
string theory, but we will deal with single-string
entropy and therefore our results do not apply directly.

\section{The microcanonical ensemble}\label{microensemble}

We want to determine statistical properties of massive string states,
in particular concerning their spatial distribution. We will use the 
{\em microcanonical} ensemble\footnote{We will discuss our preference
  for the microcanonical ensemble over the canonical one in section
  \ref{numberstatessection}, footnote \ref{micversuscan}.}. Let us
discuss for a moment the 
definition of microcanonical ensembles on more
general ground before specifying our case.

Ensembles are defined by 
density matrices: in usual statistical mechanics
the microcanonical one has the
form\footnote{\label{operatorialdelta} The expressions for the density
  matrix are 
  meaningful when applied to the states of a system; with that understanding
  our notation with Dirac's delta functions is clear.}
 \beq 
 \rho_E = a_E\delta(E-\hat H)
 \eeq
where $\hat H$ is the Hamiltonian\footnote{From now on a $\hat ~$ will
  distinguish an 
  operator from its value(s). At this moment we
consider discrete Hamiltonian spectrum.} of the system
and $a_E$ ensures the 
normalization of the density matrix   
  \beq
  \text{tr}[\rho_E]=1
  \eeq
when traced over the states.

We can try to modify the traditional microcanonical ensemble,
fixing the value of other observables, in order to investigate different
statistical properties of the system. Considering a discrete observable 
with associated operator $\hat Q$, we can define the density
matrix
 \beq
 \rho_{E, Q}= a_{E, Q} \delta(E-\hat H)\delta(Q-\hat Q).
 \eeq
If $Q$ represents an observable with continuous spectrum, we need
to specify a 
small interval $\delta Q$ (uncertainty) around the value of  the
observable we are
interested in\footnote{This applies of course also to the
  energy, if that is the case.}, and define
 \be
 \rho_{E, Q, \delta Q} & = & a_{E, Q, \delta Q} \, \delta(E-\hat H)\,
    \big(\theta(Q+\delta Q-\hat Q)-\theta(Q-\hat Q) \big) \nonumber \\
     & = & a_{E, Q, \delta Q} \, \delta(E-\hat H) \, 
       \int_{Q}^{Q+\delta Q}\delta(Q-\hat Q).
 \ee
We will let $\delta Q \to 0$, so that we can
write
 \beq
 \rho_{E, Q}  = a_{E, Q} \, \delta(E-\hat H)\delta(Q-\hat Q).
 \eeq
The partition function 
 \be \label{numberstatesmicro}
  G(E, Q) & = & \text{tr}[\delta(E-\hat H)\delta(Q-\hat Q)] \nonumber \\
          & = & 
   \sum_\phi \langle\phi|\delta(E-\hat H)\delta(Q-\hat Q)|\phi\rangle
 \ee 
gives the number of states having the values $E, \, Q$ for the chosen
observables\footnote{The number of
microstates $i$ having values $E$ for the energy, 
and $Q < Q_i < Q+\delta Q$ in the case of continuous observables.}. 
It is, therefore:
 \beq
 a_{E, Q} = G(E, Q)^{-1}
 \eeq

In our case we are interested in the partition function for a
certain microcanonical ensemble, defined by fixing the values 
\begin{itemize}
 \item of the (squared) {\em mass} \footnote{We define $\hat N_{L} \equiv
   \sum_{n=1}^\infty \hat\alpha_{-n}\hat\alpha_{n} -1$ as the
   left-moving level number operator, whose value is fixed
   once we fix the mass and charges. The right-moving one is similarly
   defined in terms of tilded oscillators. Note that here we write the
 tree-level mass.} $M_0^2=N_{R(L)}+Q^2_{R(L)}$     
 \item of the (squared) {\em size} $R^2$ of the strings\footnote{As we
   will see in section \ref{measuresetup}, the squared size is more
   easily defined in string theory.} 
 \item of the Neveu-Schwarz charges $Q_{L, R}$ (see (\ref{NScharges})).
 \end{itemize}

Then, the partition function can be written in the form:
 \beq \label{complpartfunct}
  G_c(N, R, Q_{L, R})  =   
\text{tr}[\delta(N_R-\hat N_R)\,\delta(N_L-\hat N_L)\,\delta(R-\hat R_s)\,\delta(Q_{L}-\hat Q_{L})\,\delta(Q_{R}-\hat Q_{R})]
 \eeq
Note that, since the squared size is more easily defined in string
theory (see section \ref{measuresetup}), we always define:
 \beq \label{defdeltaRs}
  \delta(R-\hat R_s) \equiv 2R \,\delta(R^2-\widehat{R^2}_s)\, .
 \eeq 
We choose not to write formal definitions of partitions
functions in terms of $\delta(R^2-\widehat{R^2}_s)$ in order to make more
readable our 
formulas. It is understood, then, that in the
following we always actually deal with the operator $\widehat{R^2}_s$,
and use (\ref{defdeltaRs}) in the formal definition of our partition function.

The apparently straightforward computation of (\ref{complpartfunct})
has its main difficulty in 
defining the string size operator $\widehat{R^2}_s$ in full
consistency with the quantization of the theory.
We will show in 
section \ref{partifuninstr} how in fact (\ref{complpartfunct}) can be actually
computed within String Theory, but first we will discuss in more
detail the issue of defining $\widehat{R^2}_s$.

\section{Measuring the size of strings}
\label{measuresetup} 

The size of a string is usually covariantly
defined in the classical
theory, as the average (squared) size
through the formula
 \beq \label{classicalaverageradius}
  \mathcal{R}_{\text{cl}}^2 = 
  {1 \ov \Delta\sigma_+\Delta\sigma_-}\int_0^{\Delta\sigma_+} \int_0^{\Delta\sigma_-}
    (X_\perp(\sigma_+, \sigma_-))^2 \quad \sigma_{\pm}=\sigma\pm\tau,
 \eeq
where
 \beq
  X^\mu_\perp(\sigma_+, \sigma_-) \equiv \tilde X^\mu-p^\mu {p \cdot\tilde X \ov p^2}
 \eeq
and 
 \beq
  \tilde X^\mu \equiv X^\mu-X^\mu_{\text{cm}}
 \eeq
where $X^\mu_{\text{cm}}$ is the center of mass motion of the
string and $p^\mu$ is its center of mass momentum. In this way
$X^\mu_\perp$ represents the projection of the 
oscillator part $\tilde X^\mu$ of
the string coordinate 
orthogonally to the center of
mass momentum of the string. $\Delta\sigma_\pm$ are the periodicities
of $\sigma_\pm$.

This classical quantity is the one that
has been used in the literature when comparing strings to black holes,
regarding size (for 
example see \cite{DamVenSelf, Cornalba, IRbh}) and this one will be
dealt with in this work.

Using the expansion  
 \beq
 \tilde{X}^\mu(\sigma, \tau) =
 \sqrt{2}\,
 \sum_{\substack{m=-\infty\\m\neq 0}}^\infty \left({\alpha^\mu_m \ov m}e^{i(\tau-\sigma)}+{\widetilde{\alpha}^\mu_m \ov m}e^{i(\tau+\sigma)}\right),
 \eeq
we find, in the string rest frame,
 \beq \label{Rclasssqparts}
 \mathcal{R}_{\text{cl}}^2 = \mathcal{R}_L^2+\mathcal{R}_R^2,
 \eeq
where
 \beq \label{classicR2L}
  \mathcal{R}_L^2 \equiv 2 \sum_{i=1}^d 
 \sum_{n=1}^\infty\left({\alpha^i_{-m}\alpha^i_{m}+\alpha^i_{-m}\alpha^i_{m} \ov n^2}\right)\, .
 \eeq 
$\mathcal{R}_R^2$ is defined as $\mathcal{R}_L^2$ with the
correspondent tilded quantities.

The seemingly most straightforward choice, at this point, would be to keep the
definition (\ref{classicalaverageradius},
    \ref{Rclasssqparts}, \ref{classicR2L}) also when quantizing the
string and promoting the $\alpha^\mu_m$'s (and $\widetilde\alpha^\mu_m$'s)
to operators $\hat\alpha^\mu_m,
\hat{~\widetilde{\alpha}^\mu_m}$. Unfortunately,
three evident issues regarding such operator are:
 \begin{itemize}
 \item its definition is gauge-dependent,
 \item the operator has a zero-order contribution
     proportional to 
     \beq
      \sum_{n=1}^\infty {1 \ov n}
     \eeq
  which needs to be interpreted and regularized (see \cite{SussSizeString}),
 \item the insertion of this operator in
       a path-integral is problematic because 
 $\widehat{\mathcal{R}^2}$
 defined in this way is not BRST invariant, in a covariant gauge
 formulation of String theory, or equivalently, in an Old Covariant
 quantization scheme, it generates unphysical
 states when applied to physical ones. 

 Instead, when using a Light-Cone gauge quantization, the presence
 in $\widehat{\mathcal{R}^2}$ of the longitudinal oscillators $\alpha^-_m =
 (p^+)^{-1}L^{\text{transverse}}_m$ leads to a complicated interacting
 theory for the transverse oscillators and poses ordering problems when
 used in composite operators, such as our desired delta function.  
 \end{itemize}
~~
It appears therefore clear that one cannot carry over consistently
in String Theory the minimal program of
a) considering formula (\ref{complpartfunct}),
b) expressing the delta function\footnote{For example through Fourier
    transform or similar.} in terms of operators and using the
    $\widehat{\mathcal{R}^2}$ 
    defined in the naive way we just illustrated, 
c) tracing over physical string states.

We will instead work only with
quantities\footnote{Namely amplitudes.} that are 
well-defined in string 
theory. From them, we will define and compute
the partition function in terms of mass, charge and size. 
The classical value (\ref{classicalaverageradius}) must obviously be
recovered in the (semi)classical limit. We will
use this as a key test in verifying the correctness 
of our definition for the operator in question (see section
\ref{numberstatessection}). 

The procedure we will follow to obtain (\ref{complpartfunct}) will
then be:
 \begin{itemize}
  \item[{\bf a})] compute the partition function\footnote{Here we again
   have 
  \beq \label{defdeltaR}
   \delta(r-\hat R)\equiv 2r\,\delta(r^2-\widehat{R^2}),
  \eeq 
as in (\ref{defdeltaRs}).} 
    \beq \label{partialpartfunc}
     G^*=\text{tr}[\delta(N_R-\hat N_R)\,\delta(N_L-\hat N_L)\,\delta(r-\hat R)\,\delta(Q_{L}-\hat Q_{L})\,\delta(Q_{R}-\hat Q_{R})]
    \eeq
 for an operator size $\hat R$, to be defined
  starting from well-defined string amplitudes and from a physical
  procedure to measure the size of objects (see section \ref{partifuninstr}).
  \item[{\bf b})] prove that the operator $\widehat{R^2}$
    defined by this procedure recovers
    the 
    classical value and form given by (\ref{classicalaverageradius},
    \ref{Rclasssqparts}, \ref{classicR2L}), when
    evaluated on the string ensemble, and that therefore 
    it is the correct string squared size operator 
    $\widehat{R^2}_s$, so that
    $G^*$ is indeed $G_c(N, R, Q_{L, R})$ as written in (\ref{complpartfunct}).
  \end{itemize}

Let us now define and obtain $G^*$.

\section{Size distribution of highly excited free string states}
\label{partifuninstr} 

To avoid cluttering of formulas, for the moment we will consider
states with zero charges; in section
\ref{stateswithcharges} we will extend 
our method to non-zero fixed charges.

Let us start by considering the on-shell string
amplitude defined as\footnote{$G_c(N)$ is the number of closed string 
   states at mass 
  level $N$. Note that
  $G_c(N)=G_o(N)^2$, where $G_o(N)$ is the number of open string
  states at level mass $N$.}
 \beq \label{borampli}
 A_{\text{closed}}={g_s^{2} \ov G_c(N)}\int d^2\,z \, 
  \sum_{\phi_{|_N}}\langle \phi|V(k', 1) V(k, z)|\phi\rangle.
 \eeq
The sum is over physical string states at fixed squared mass $N$
extended in the large dimensions.
The on-shell vertex operators\footnote{Our vertex operators do not
  have the usual string coupling factor carried by the string
  vertex operators. This is because for clarity we have decided to
  explicitly show all string coupling factors in front of our
  amplitudes in this paper.}
 \beq \label{gravvertex}
 V(k, z)= {2 \ov \alpha'} e^{i k\cdot X} 
          (\partial X^\mu-{i \ov 2} \psi^\mu k\cdot\psi)
          (\bar\partial X^\nu-{i \ov 2}
              \tilde\psi^\nu k\cdot\tilde\psi)\, \xi_{\mu\nu}, \quad
              k^2=k^\mu \xi_{\mu\nu}=0
 \eeq
represent gravitons (actually a
superposition of graviton, dilaton and Kalb-Ramond field, in the
following indicated by the letter $b$).  
Here, $X, \psi$ are, respectively, the
space-time string bosonic and fermionic coordinates.

Our goal will be computing (\ref{borampli}) and show how from it we
can obtain (\ref{partialpartfunc}).

The amplitude can be conveniently re-written as a trace inserting the
density matrices (mass projectors) $\rho_N, \tilde \rho_N$,
where
 \beq \label{massproject}
 \rho_N = {1 \ov \sqrt{G_c(N)}}\, \delta(N -\hat N_L)=
  {1 \ov \sqrt{G_c(N)}}{1 \ov 2\pi i}\oint {dw \ov w^{N+1}}w^{\hat N_L} \,.
 \eeq
Note that $N_R=N_L=N$ for the case we
 are considering.

We obtain
 \beq \label{treeaveoneloop}
 A_{\text{closed}}=g_s^{2}
  \int d^2\,z \, tr[V(k', 1) V(k, z)\rho_N\tilde \rho_N] .
 \eeq 
In the form (\ref{treeaveoneloop}), the
amplitude can be computed as a one-loop two point amplitude for the
probes $b$ (projected on 
``initial'' states of mass-level $N$ and without integrating over the
zero modes). This ensures that  only physical states enter in the
trace and therefore that this on-shell amplitude is
well-defined in string theory.

\begin{figure}[t]
 \begin{center}
\includegraphics[height=4cm]{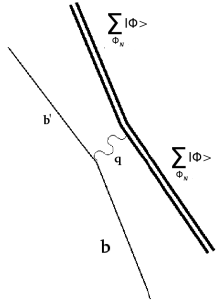}
\end{center}
\caption{{\small \sl{Scattering process $b+\sum_N\phi \to b'+\sum_N\phi$. At low
      momentum transfer the process is dominated by the massless
      channel, which is represented here. }}} 
 \label{bornapprox}
\end{figure}

The process accounted for by (\ref{borampli}) is the one depicted in figure 
\ref{bornapprox}, that is the scattering of $b$ off the average string state
at mass-level $N$.
Note that what we are computing here is an amplitude, not an {\em
  inclusive} cross 
section, as had instead been done in \cite{AmRus}
for the bosonic string and later in \cite{CIR} for the
superstring.

We will study this amplitude for elastic scattering (when $q^2= \vec q^{~2}$),
low momentum transfer $(k+k')^2=q^2=
\vec q^{~2}\to 0$ and energetic probes and massive
states\footnote{This corresponds to the Regge limit of 
our amplitude.} such
that $-k\cdot p \gg 1$ (see section \ref{correctionsfact}). In this
limit the amplitude can be written as (proof will be given in section
\ref{correctionsfact})
 \beq \label{factorizformfact}
 A_{\text{closed}} \sim {(E\sqrt{N})^2 \ov q^2}F_{b}(q^2) F_{N}(q^2)
 \eeq
and $F_i(q^2)$ is the {\em form
 factor}
of the state $i$. In our formula we used the symbol $F_N$ because
we are interested in a microcanonical ensemble at
  fixed mass level number $N=M^2$. 

We also know that the form factor is the Fourier transform of certain
spatial distributions  of the string. For example, if the probe $b$ is a gauge
field excitation then we recover the relative charge spatial distribution
$\mu_Q(\vec r)$, if  
$b$ is a graviton (as in our case), we recover the mass spatial
distribution $\mu_N(\vec r)$.

We therefore write
 \beq \label{formfactmassdistr}
  F_N(q^2)  =  \int dq e^{i \vec q\cdot \vec x}\mu_N(\vec r)
 \eeq
and we recall that (we consider here the case of spherical symmetry, which
 will occur in our case, $\int d\Omega_d$ is the angular integral)
 \be \label{stringmassRdistr}
  \int d\Omega_d\,\mu_N(\vec r) & = & \Omega_d\,r^{-(d-1)} \mu_N(r)
   \\
  & = & 
  r^{-(d-1)}\, G^{-1}_c(N)\, tr[\delta(r-\hat R)\delta(N-\hat N_R)\,\delta(N-\hat N_L)]
  \nonumber
 \ee
using (\ref{massproject}, \ref{treeaveoneloop}) and the
definition of mass spatial distribution in a quantum theory for states
with mass $M=\sqrt{N}$.

At this point we define
 \be \label{interpartfunc}
 G^* & = & \text{tr}[\delta(N-\hat N_R)\,\delta(N-\hat N_L)\,\delta(r-\hat R)]
  \nonumber \\
   & = & G_c(N) \Omega_d\,\mu_N(r) \,, 
 \ee
where we have used (\ref{formfactmassdistr}, \ref{stringmassRdistr}).

The operator $\widehat{R^2}$ has been therefore defined in a
operational way. What we are doing here really represents the
correct method for physically defining and measuring the size of an
object: through scattering processes. 
The average
value of $\widehat{R^2}$ is given by
 \beq \label{avquantdef}
 \langle\widehat{R^2} \rangle = -2 d\, \partial_{q^2} F_N(q^2)|_{q^2=0}\, .
 \eeq

In order to show that $G^*$ is indeed the
looked-for partition function (\ref{complpartfunct}), we have to demonstrate
that the operator $\widehat{R^2}$ defined in this way
really measures the size of a string, 
that is to say, it recovers the correct classical value and form
(\ref{classicalaverageradius},
    \ref{Rclasssqparts}, \ref{classicR2L}) in the (semi)classical
    limit. This can 
be done using 
(\ref{avquantdef}), which enables us
to compare the average value of our operator
$\widehat{R^2}$ with the average size
of the string computed classically with (\ref{classicalaverageradius}). 
We will prove this in
section \ref{numberstatessection}\footnote{Proving this correspondence
  will actually require that we 
formally use the canonical ensemble, but that will be straightforward to obtain
from our formulas, as we will show.}.

An additional new result that we obtain here is the
computation of the form factor for a string ensemble at fixed squared mass.
Note that this has not been previously done in the literature\footnote{
Our formulas, and therefore the result that we will obtain,
are indeed very different from the
  ones in \cite{ManesFF1}. 
In our case the
  interpretation of $F_N(q^2)$ as 
  a form factor is justified, according to scattering theory, 
  whereas in \cite{ManesFF1} it
  could not be accepted because the form factor of an object cannot be
  obtained from the {\em semi-inclusive} cross-section of its decay.
In fact this was acknowledged by the author of \cite{ManesFF1} himself
in the successive paper \cite{ManesFF2} (see section 3 there), where 
a different and more correct 
  interpretation of the results was proposed.}.

In the following, we compute (\ref{borampli}) both for the bosonic
string theory and for the superstring. 
The computations are in fact very similar. We
therefore discuss  
first and at length the superstring, and later address the bosonic.

\subsection{States with no charge} \label{sizesec} 

As we said, we concentrate at first on the case where the 
string states in our ensemble carry no charge. The formulas are in fact neater 
and all the steps can be discussed in a clearer way. In section
\ref{stateswithcharges}, 
instead, we will consider non-zero charges of the
Neveu-Schwarz type.

\subsubsection{The String Spatial Distribution}\label{SpatialDistribution}

\paragraph{The superstring.}
~~

In computing (\ref{borampli}, \ref{treeaveoneloop}),
we will make use of the relation\footnote{Here we have explicitly
  reinstate $\alpha'$ 
  in order to present clearly the formula. Remember
  that in the computations we will always set $\alpha'=4$.}
\cite{KLT}
 \beq
 A_c(1234; \alpha', g_s)={\pi i g_s^2\alpha' \ov g_o^4}
 \sin(\pi\alpha't) A_o(s, t; {\alpha' \ov 4}, g_o) \tilde A_o(t, u; {\alpha' \ov 4}, g_o).
 \eeq
where  $s, t, u$ are Mandelstam variables. 

The amplitude that we will compute is therefore\footnote{This
  well-defined string
  amplitude can be computed both in the covariant and in the
  light-cone gauge, at each one's own convenience. We present here the
results, which are independent of the chosen gauge.}
 \beq \label{borampliop}
 A_o(s, t; 1, g_o)=  g_o^2 \int dy
   \,\text{tr}[V_{\text{open}}(k', 1)V_{\text{open}}(k, y)\rho_N]
 \eeq
with, given (\ref{gravvertex}), 
 \beq
 V_{\text{open}}(k, y) = {e^{i k\cdot X(y)} \ov \sqrt{ 2\alpha'}}  
          \left(i y \xi\cdot\partial_y X(y)+2\alpha'  k\cdot\psi(y) \xi\cdot \psi(y)\right)
  \qquad k^2=k\cdot \xi=0\,.
 \eeq
We will consider the limits $t\equiv -q^2=-(k+k')^2 \to 0, \,
-k^0\sqrt{N} \gg 1$. 
The leading term of the amplitude in this limit can be calculated using 
the OPE (see \cite{ManesFF1})
 \beq \label{OPE}
 V_{\text{open}}(k', 1) V_{\text{open}}(k, y) \underset{y\to 1}{\sim} g_o^2
2\xi\cdot\xi' \text{{\small $(1-q^2)\,(1-y)$}}^{\text{{\tiny $2 k'\!\!\cdot \!\!k\!\!-\!\!2$}}}y^{\text{{\tiny $2 k\!\!\cdot \!\!\hat p$}}}
      e^{i q\cdot \hat X_O(1)} e^{iq\cdot \hat x}
 \eeq
where $\hat X_O$ indicates the oscillator part of $X$.
Note that, by performing this OPE, we are actually using
the property of factorization of (string) amplitudes. This will turn out to
be very useful:
factorizing two external legs of an amplitude, the momentum square
$q^2$ flowing 
along the connecting propagator is a continuous variable, allowing 
analytic continuation.

The
amplitude can be written as
 \beq
 A_o =g_o^2 A_o^{\text{zero modes}}A_o^{\text{oscillators}}
 \eeq
By writing $y = e^{-\epsilon}$ with $\epsilon\to 0 $, we find the 
result\footnote{We need to perform the same analytical continuation as
 for the Veneziano 
amplitude, as usual in these representation of the string amplitudes.}
 \be \label{integralGammaSuper}
 A_o^{\text{zero modes}} & = &
     -\int d\epsilon \,\epsilon^{q^2-2}e^{-\epsilon(2 k\cdot p+1)}\, (1-q^2)
      \\
   & & \underset{q^2 \to 0}{\sim} {(2\sqrt{N}E)^2 \ov q^2} \sqrt{F_b(q^2, E)} \, (2\sqrt{N})^{-q^2}.
 \ee
where we have defined 
 \beq \label{masslessformfactor}
  F_b(q^2, E) \equiv e^{-2q^2\ln(2E)}
 \eeq 
and $E\equiv k^0$.

Therefore,
 \beq
 A_c(1234; 4, g_s)\sim \pi^2 i g_s^2 
  {(2\sqrt{N}E)^4 \ov q^2} F_b(q^2, E) (2\sqrt{N})^{-2q^2} 
  A_o^{\text{oscillators}} \tilde A_o^{\text{oscillators}}
 \eeq
where
 \beq \label{Aopenoscill}
  A_o^{\text{oscillators}} = \text{tr}[e^{i q\cdot X_O(1)} \rho_{N}]
 \eeq
and we have expanded $\sin(-\pi t)\sim -\pi t \sim \pi q^2$. 

According to the results and the discussion in \cite{SussBlStrCompl},
we identify   
$F_b(q^2, E)$ with the form factor for the probe $b$\footnote{Note
  that \cite{SussBlStrCompl} sets $\alpha'=2$, we use $\alpha'=4$}.

It is now straightforward to read the form factor for the target
average state at squared mass $N$:
 \be \label{formfactorN}
 F_N(q^2) & = & (4N)^{-q^2} A_o^{\text{oscillators}} \tilde A_o^{\text{oscillators}} 
 \\ 
 & = &  {4N^{-q^2} \ov G_c(N)} {1 \ov (2\pi i)^2} \oint {dw \ov w^{N+1}} 
     \oint {d\tilde w \ov \tilde w^{N+1}}
   {g(w)g(\tilde w) \ov (f(w)f(\tilde w))^{d-1}} \nonumber \\
 & & ~~~~~~~~~~~~~~~~~~~~~~~~~~
   \times e^{-2q^2\sum_{n=1}^\infty{w^n \ov n(1-w^n)}+{\tilde w^n \ov n(1-\tilde w^n)}}
 \nonumber 
 \ee
where 
 \begin{align} \label{partfunc}
  f(w)= \prod_{n=1}^\infty (1-w^n) \qquad 
   g(w)= {1 \ov \sqrt{w}}g_3(w)^{d-1}-{1 \ov \sqrt{w}}g_4(w)^{d-1}+ g_2(w)^{d-1}
   \, \nonumber \\
 g_3(w)=\prod_{r={1\ov 2}}^\infty(1+w^r) \quad
 g_4(w)=\prod_{r={1\ov 2}}^\infty(1-w^r) \quad
 g_2(w)=\prod_{r=0}^\infty(1+w^r) \ .
 \end{align}
We compute the loop-integrals by saddle point approximation for large $N$,
finding
 \beq \label{saddlepointsize} 
 \ln(w) \sim - {\pi \ov \sqrt{N}}\sqrt{{d-1 \ov 4}-{q^2 \ov 3}}
 \eeq
and similarly for $\tilde w$.

Therefore, in the elastic limit, for small $\vec q²$,
 \be
 F_N(\vec q^{~2}) & \underset{N\to\infty}{\sim} & 
  {e^{4\pi\sqrt{N}\sqrt{{d-1 \ov 4}-{\vec q^{~2} \ov 3}}} \ov G_c(N)}\pi^{d}
   \left({(d-1) \ov 4 }-{\vec q^{~2} \ov 3}\right)^{{d \ov 2}}
   N^{-{d+2 \ov 2}-\vec q^2}
  \nonumber \\
   & \underset{\substack{\vec q^{~2}\to 0 \\ \\ N \to \infty}}{\sim} & 
   e^{-{4\pi \ov 3} \sqrt{{N \ov d-1}} \vec q^{~2}}
 \ee
where in the last line we have simplified the result with 
 \beq
   G_c(N) \sim e^{2\pi\sqrt{N\,(d-1)}}\pi^{d}
      \left({d-1 \ov 4}\right)^{{d \ov 2}} N^{-{d+2 \ov 2}}.
 \eeq
Finally, the mass distribution is
 \beq \label{sizedistribution}
 \mu_N(\vec r) = 
  {1 \ov (2\pi)^d} \int d^dq e^{i \vec q \cdot \vec x} F_N(\vec q^{~2})=
 \left({3 \ov 16\pi^2} \sqrt{{d-1\ov N}} \right)^{{d \ov 2}}
 e^{-{3 \ov 16 \pi}\sqrt{{d-1\ov N}} \vec{r}^2},
 \eeq
where $\vec r^2= \vec x^2$.
\vskip 0.4cm

\paragraph{The bosonic string.}
~~
\nopagebreak

The case of the bosonic string follows the same steps. 
A few things are different:
 \begin{itemize}
 \item the vertex operator for the probe now is
    \beq \label{gravvertexbosonic}
      V_{\text{open}}(k, y) = {1 \ov \sqrt{ 2\alpha'}} e^{i k\cdot X(y)} 
          \left(i y \xi\cdot\partial_y X(y)\right)
      \qquad k^2=k\cdot \xi=0 
    \eeq
 \item due to the absence of fermionic excitations, the 
   number of closed string states at fixed large mass squared $N$ is
   \beq
     G_c(N) \sim e^{4\pi\sqrt{N}\sqrt{{d-1 \ov 6}}}\pi^{d}
    \left({(d-1) \ov 6 }\right)^{{d \ov 2}} N^{-{d+2 \ov 2}};
   \eeq
  \item the integral (\ref{integralGammaSuper}) becomes now
   \beq \label{integralGammaBosonic}
    A_o^{\text{zero modes}}  = 
     -\int d\epsilon \,\epsilon^{q^2-2}e^{-\epsilon(2 k\cdot p+1)}\,.
   \eeq
  \end{itemize}
Namely, we see that the integral would be divergent also for $q^2=1$, 
corresponding to the exchange of a tachyon. But we are considering the limit 
$q^2 \to 0$, picking out the graviton pole, so that
 \beq 
 A_o^{\text{zero modes}}  
 \underset{q^2 \to 0}{\sim} {(2\sqrt{N}E)^2 \ov q^2} \sqrt{F_b(q^2, E)} \, (2\sqrt{N})^{-q^2}
 \eeq
as for the superstring. 
Therefore we obtain
 \begin{itemize}
  \item form factor
     \beq
       F_N(\vec q^{~2}) \underset{{\vec q^{\,2}\ov N}\to 0}{\sim}  
         e^{-2\pi \sqrt{{2 \,N \ov 3\, (d-1)}} \vec q^{~2}},
     \eeq
  \item mass distribution
     \beq \label{sizedistributionbos}
      \mu_N(\vec r) = 
       \left({1 \ov 8 \pi^2}\sqrt{{3\,(d-1) \ov 2N}} \right)^{{d \ov 2}}
       e^{-{1 \ov 8 \pi}\sqrt{{3\,(d-1) \ov 2N}} \vec{r}^2}.
     \eeq
 \end{itemize}
The number of extended spatial dimensions now can go up to $d=25$, not
only up to 9 as for the superstring.

\subsubsection{Corrections}\label{correctionsfact}

We show here how the lowest terms in the OPE
for $y \to 1$ in (\ref{OPE}), indeed dominate the amplitude (\ref{borampliop})
and the result is safe against possible corrections 
in the considered kinetic and mass range. In particular, within this
range, formula (\ref{factorizformfact}) appears correct and allows the
definition of form factors. 

\paragraph{The superstring.}
~~
\nopagebreak

Without any approximations, the amplitude (\ref{borampliop}) is given
by
 \be
 A_o(s, q^2;1, g_o) & \sim & \sum_{s=2}^4{g_o^2 \ov G_o(N)}\!\int\!\!d\epsilon 
  {1 \ov 2\pi i}\oint\!\!{dw \ov w^{N+1}}
  {g(w) \ov f(w)^{d-1}}\,\,
  e^{-\epsilon (2k\cdot p+1)}\,\,\psi(\epsilon, w)^{q^2}\,  \nonumber \\
  & & ~~~~~~~~~~~~~~~~~~  \times 
 \left[-2\partial^2_\epsilon \ln{(\psi(\epsilon, w))} + 
       \chi_s(\epsilon, w)\right]
 \ee
with
 \be \label{logterm}
 \psi(\epsilon, w) & = & 
  (1-e^{-\epsilon})
 \prod_{n=1}^\infty e^{-q^2 {w^n \ov n(1-w^n)} (e^{n\epsilon}+e^{-n\epsilon})} \\
 \partial^2_\epsilon\ln{(\psi(\epsilon, w))} & = & 
  \sum_{n=1}^\infty n e^{-\epsilon n} +
 \sum_{n=1}^\infty {n w^n \ov (1-w^n)}(e^{n\epsilon}+e^{-n\epsilon})  \\
 \chi_s(\epsilon, w) & = & 2q^2 
 {\theta_s(\epsilon)^2 \theta_1'(0)^2 \ov\theta_1(\epsilon)^2\theta_s(0)^2} ,
 \ee
where we have written $y=e^{-\epsilon}$. Note that $\theta_s(z) \equiv
\theta_s({z \ov 2\pi i}, {\ln(w) \ov 2\pi i})$ in the usual notation,
where the $\theta_s$'s are the 
Theta functions.

Expand for $\epsilon \to 0$: 
 \be
 I_\epsilon & \sim & \int d\epsilon e^{-\epsilon\,(2k\cdot p+1)} \epsilon^{q^2-2}
  e^{-2q^2\sum_n {w^n \ov n (1-w^n)}}
  \left(1-q^2+O\left({w\epsilon^2 \ov (1-w)}\right)\right)
  \nonumber \\
  & \sim & (2k\cdot p)^{-q^2+2}\Gamma(q^2)\left(1+O({\sqrt{N}\ov (k\cdot p)^2})\right). 
 \ee
In the limit $-k\cdot p =E\sqrt{N} \gg 1$ that we have been
considering ($E$ probe energy, 
 $\sqrt{N}$ tree-level mass for the massive 
 state) our results
 appear to be valid.

\vskip 0.4cm
\paragraph{The bosonic string.}
~~
\nopagebreak

The bosonic string case is similar to the superstring one: it can be
quickly obtained eliminating  
from the formulas above the term $\chi(\epsilon, w)$ and substituting 1
to $g(w)$, which leads to the
result
 \beq
 I_\epsilon \sim  (2k\cdot p)^{-q^2+2}\Gamma(q^2-1)\left(1+O({\sqrt{N}\ov (k\cdot p)^2}) \right). 
 \eeq
showing again the validity of our expansion for $E\sqrt{N} \gg 1$.

\subsubsection{Identifying the string size operator and recovering the partition function of string states of a given mass and size}
\label{numberstatessection}

In this section we finally show that the operator $\hat R$ 
in (\ref{interpartfunc}) does indeed recover the classical value and form
given by (\ref{classicalaverageradius}) in the (semi)classical limit.

Note that we have never
written or supposed and expression for $\widehat{R^2}$ in term of
string oscillators whatsoever. All we have used (and will consistently
use here as 
well) is a procedural definition of the quantum operator
through scattering. This we will compare with the classical formula, which is
written in terms of string coordinates.

Let us start with the quantum computation. The average value of
the quantum operator defined through scattering in the previous
section(s) can be easily computed from 
 \beq \label{avRsqinterquant}
  \langle \widehat{R^2} \rangle =  -2 d \,\partial_{q^2} F_N(q^2)|_{q^2=0}\,.
 \eeq
In order to compare this to the classical result 
(\ref{classicalaverageradius}, \ref{Rclasssqparts}, \ref{classicR2L}), it is
convenient to adopt a canonical ensemble formalism, rather than a
microcanonical\footnote{\label{micversuscan}Let us comment at this point on
  advantages and 
  disadvantages of the 
canonical formalism.
The canonical ensemble is more convenient than the microcanonical in
analyzing values of operators because it somehow preserves the structure of
the operator itself (in this case the sum we see in (\ref{classicR2L})),
whereas the microcanonical 
ensemble would give us just a number, corresponding basically to the
sum already performed. In closed string theory, however, the canonical
formalism presents a fundamental problem: closed strings must obey the {\em
  level matching condition} which sets
 \beq \label{levelmatch}
   L_0|\phi\rangle= \tilde L_0 |\phi\rangle
 \eeq
where $L_0$ ($\tilde L_0$) are the Virasoro operators which implement
the mass constraint. This, in our case, reduces to
 \beq
  N_L = N_R = N 
 \eeq
for {\em every single string state}, where $N_R$ depends on tilded
oscillators and $N_L$ on not-tilded 
ones. It appears evident that the naive canonical ensemble cannot
ensure that (\ref{levelmatch}) is satisfied for every
state and therefore
it is not guaranteed that we are tracing only over physical states.}.
Formally, this can be done simply by not performing the
integral ${1 \ov 2\pi i}\oint dw\, w^{-N-1}$
in the formulas of the preceding section and setting
$w=\tilde w=e^{-\beta}$, so that
 \beq
  Z_{\text{micro}}=tr[\delta(N_R-\hat N_R)\,\delta(N_L-\hat N_L)] \to
  Z_{\text{can}}=tr[e^{-\beta \hat N_R}\,e^{-\beta \hat N_L}]
 \eeq
with $\beta$ related to the squared mass by $-\partial_\beta
\ln(Z_{\text{can}})= \langle \hat N \rangle = N$.

From (\ref{avRsqinterquant}) and the results in the previous section
(see (\ref{formfactorN})),
we then find 
 \beq \label{averRsqinterm}
  \langle \hat R^2 \rangle =  4d\sum_{n=1}^\infty\left({w^n \ov n(1-w^n)}+{\tilde w^n \ov n(1-\tilde w^n)}\right).
 \eeq
Here we have chosen to write $w, \tilde w$
instead of their common value $e^{-\beta}$ for an easier comparison
with previous formulas in the paper.

\vskip 0.2cm
Let us now turn to the classical value given by 
(\ref{Rclasssqparts}, \ref{classicR2L}). This has to be compared with
the quantum result in the (semi)classical limit. For a discussion about
the definition of the
classical limit of quantum mechanics see \cite{Ballentine}. 
In the simplest formulation average values of
quantum operators are put in relation with classical values. This
means that in the limit 
we can write\footnote{Where not explicitly
  written, no sum over the index $i$ is performed in the
  following.}
 \beq \label{averalpcan}
  \alpha^i_{-m}\alpha^i_{m} \leftrightarrow 
   \langle \hat \alpha^i_{-m}\hat \alpha^i_{m} \rangle \,.
 \eeq
Now,
neglecting the motion of the center of mass, states of a closed
(therefore periodic) string are standing waves of various integer
frequencies $n$. We will use a semiclassical limit, instead than a
fully classical one, in order to preserve the discreteness of energies of
standing waves. 
The true classical
behavior can then be obtained at high temperature (small $\beta$), large
quantum numbers.

The total energy of the string is
 \beq
  N = N_L+ N_R
 \eeq
 \beq
  N_L=  
  \sum_{n \geq 1, i}\alpha^i_{-n} \alpha^i_{n} 
  \equiv \sum_{n \geq 1, i} n N^i_{L, n}
  \eeq
  \beq
  N_R=  
   \sum_{n \geq 1, i}^{+\infty} \tilde \alpha^i_{-n} \tilde \alpha^i_{n}
   \equiv \sum_{n\geq 1, i} n N^i_{R, n}
 \eeq
As is well known, then, the occupation number
relative to a single wave energy level $n$ in the canonical ensemble
is given by $N^i_n=(e^{\beta n}-1)^{-1}$.
By substituting this in 
(\ref{Rclasssqparts}, \ref{classicR2L}), always in the
semiclassical limit (\ref{averalpcan}), we obtain
 \beq \label{clasRsqonensem} 
 \mathcal{R}_{\text{cl}}^2 = 4d\sum_{n=1}^\infty\left({w^n \ov n(1-w^n)}+{\tilde w^n \ov n(1-\tilde w^n)}\right) \, .
 \eeq
We clearly see that in this limit
 \beq
  \langle \widehat{R^2} \rangle =
  \mathcal{R}_{\text{cl}}^2\, .
 \eeq
The quantum operator will actually also have a zero point contribution
from the normal ordering. This latter 
gives origin to the sub-leading factor $N^{-q^2}$ in
(\ref{formfactorN}) 
which is negligible for $N \to \infty$. 

As we already stressed, the
computations of the quantum average (\ref{averRsqinterm})
and of the classical value (\ref{Rclasssqparts},
\ref{clasRsqonensem}) 
are completely independent: the former evaluates a quantum operator
defined in a consistent procedural way through scattering 
with no reference to an
expression in terms of string oscillators, the latter uses a
classically well defined formula in terms of oscillators in the
semiclassical limit.

Having shown that the average values of $\widehat R^2$
correctly gives the classical value 
(\ref{classicalaverageradius}) of the
string size in the semiclassical limit,
we can identify
(\ref{interpartfunc}) with the the partition function $G_c(N, R)$ for string
states with mass 
level $N$ and size $R$ and write
 \beq
  G^*=G_c(N, R) \, .
 \eeq

\subsubsection{Partition function for strings at fixed mass and size}\label{sizesecfinform}

Using now formulas (\ref{formfactmassdistr},
\ref{stringmassRdistr}, \ref{interpartfunc}),  
we obtain
\begin{itemize}
 \item for the {\bf superstring}
 \begin{itemize}
  \item the partition function for closed string states with fixed $R, N$
   \beq \label{numberfixedNR}
    G_c(N, R)\!\! =  \!\!{2 \ov \Gamma({d \ov 2})}
     \left({3 \sqrt{d-1} \ov 16\pi^2 \sqrt{N}} \right)^{\!{d \ov 2}} \left({R \ov \sqrt{N}}\right)^{\!d\!-\!1} 
   {e^{\pi\sqrt{d-1}\left(2\sqrt{N}-{3 \ov 16 \pi^2\sqrt{N}} R^2\right)}\ov N^{3 \ov 2}}
   \eeq
  \item and the entropy
   \be
     S & = & \ln(G_c(\sqrt{N}, R)) \nonumber \\
      & \sim & 2\pi \sqrt{N}\sqrt{d-1}-{3\sqrt{d-1} \ov 16\sqrt{N} \pi}  R^2 + 
       \ln\left({R^{d-1} \ov \sqrt{N}^{{3\ov 4}d+1}} \right)
   \ee
  \end{itemize}
 \item for the {\bf bosonic string}
 \begin{itemize}
  \item the  partition function for closed string states with fixed $R, N$
   \beq \label{numberfixedNRbos}
    G_c(N, R)\!\! = \!\!{2 \ov \Gamma({d \ov 2})}
     \left({\sqrt{3 (d-1)} \ov 8\sqrt{2}\pi^2 \sqrt{N}} \right)^{\!\!{d \ov 2}}\!\!\!\! \left({R \ov \sqrt{N}}\right)^{\!\!d\!-\!1} 
  \!\! {e^{\pi\sqrt{d-1}\left(\sqrt{{8N \ov 3}}-{\sqrt{3} \ov 8 \sqrt{2}\, \pi^2\, \sqrt{N}}  R^2\right)}\ov N^{3 \ov 2}}
   \eeq
  \item and the entropy
   \be
     S & = & \ln(G_c(N, R)) \nonumber \\
      & \sim & 4\pi \sqrt{N}\sqrt{{d-1\ov 6}}-
      {\sqrt{3\,(d-1)} \ov 8\sqrt{2}\,\sqrt{N}\, \pi}  R^2 +
      \ln\left({R^{d-1} \ov \sqrt{N}^{{3\ov 4}d+1}} \right).
   \ee
  \end{itemize}
\end{itemize}

By maximizing the
entropy with respect to $R$ at fixed $N$, we find that the majority
of string states have the following favored
values of the size:
 \begin{itemize}
 \item for the superstring:
   \beq
    R^2_{\text{max number}} = 
       {8\pi \,\sqrt{d-1}  \ov 3} \sqrt{N}
   \eeq
 \item for the bosonic string 
     \beq
      R^2_{\text{max number}} =   
       {4\sqrt{2} \ov 3}\,\pi \,\sqrt{d-1}  \sqrt{N},
     \eeq
 \end{itemize}
The average radius is instead: 
 \begin{itemize}
 \item for the superstring:
   \beq
     R^2_\text{average} =  {8\pi \,d  \ov 3} \sqrt{{N \ov d-1}}
   \eeq
 \item for the bosonic string 
     \beq
       R^2_\text{average} =   
       4\sqrt{2}\,\pi \,d  \sqrt{{N \ov 3\,(d-1)}},
     \eeq
 \end{itemize}
which are in agreement with results (obtained with various
approximations) in the literature.

\vskip 0.4cm

A few remarks are useful at this point. First, we can appreciate in
our derivation,
the importance of the factorization property of (string) amplitudes:
as we already mentioned,
factorizing two external legs of an amplitude, the momentum square
$q^2$ flowing 
along the connecting propagator is a continuous variable, allowing 
analytic continuation. Therefore (\ref{partialpartfunc}) is
computable, using formula (\ref{interpartfunc}), in a perfectly
consistent way within string theory\footnote{Starting form the
  well-defined on-shell amplitude (\ref{borampli}).}, although the
presence of an 
off-shell insertion $\delta(R^2-\widehat{R^2}_s)$. The realization of
this point, is 
the key technical achievement that allows the computation of the quantum
partition function.

We could
also wonder whether our result depends on
the ordering of the two delta insertions in (\ref{partialpartfunc}),
since we do not expect $\widehat{R^2}_s$ and $\hat N_{R(L)}$ to commute. 
In any case, we can be reassured by the fact that obviously  
$\delta(R^2-\widehat{R^2}_s) \,\delta(N-\hat
N_{R(L)})$ and $\delta(N-\hat N_{R(L)}) \,\delta(R^2-\widehat{R^2}_s)$ 
yield the same result when traced over, and,
furthermore, we are working with very massive
string states,  
for which it is also reasonable to take a semi-classical limit.

As a final remark, we can also see that the quantum
computation we have performed has clarified the link with the random
walk approximation that has been used in the past. Indeed, in the
random walk picture a string of mass $M$ is described as a Gaussian of
width proportional to $\sqrt{M}$. The mass
distributions we have obtained in formulas (\ref{sizedistribution},
\ref{sizedistributionbos}) for the bosonic string and the superstring
have precisely that form. This compares also to the discussion in 
\cite{HorPolSelf}, section 2.

\subsection{States carrying Neveu-Schwarz charges} \label{stateswithcharges}

In this section we consider states carrying non-zero
Neveu-Schwarz type charges. The implementation of the relative delta functions
in (\ref{complpartfunct}, \ref{partialpartfunc}) is easily achieved
by fixing the Kaluza-Klein and winding mode numbers for the states in
the ensemble. We therefore report here the notation and the results.

\subsubsection{Non-BPS states}\label{nonBPSpartfunc}

The results obtained in the previous sections can be extended to
ensembles of string states carrying Neveu-Schwarz charges $Q_R, Q_L$. 
We have to distinguish states according to their mass and their
winding and Kaluza-Klein mode numbers $(m^i, n^i)$, such that:
 \be \label{NScharges} 
 Q^i_{L, R} & = & \left({n^i \ov r^i}\pm {m^i r^i \ov 4}\right) \\
 Q^2_{L, R} & = & \sum_i Q_{L, R}^{i\,2},
 \ee
where $r^i$ is the radius\footnote{Recall that we set $\alpha'=4$
  and express everything in units of $\alpha'$.} of
 compactification in the $i$-th
 compactified direction.

The mass-shell condition and the Virasoro constraint 
$(L_0-\tilde L_0)|\phi\rangle=0$ read
 \be
 M^2 & = & Q_L^2 +N_L \\
     & = & Q_R^2 +N_R \\
 N_L-N_R & = & -\sum_i n^im^i.
 \ee
where $L, R$ indicate respectively the left- and right-moving sectors.

We define our microcanonical system by fixing charge and squared mass,
which implies fixing the values $N_L, N_R$ of the
operators
$ \hat N_L,\, \hat N_R$. We consider large $N_L, N_R$.

Then, defining
 \beq
 \mathcal{N}=\sqrt{N_L}+\sqrt{N_R}\, ,
 \eeq
\begin{itemize}
 \item for the {\bf superstring}
 \begin{itemize}
  \item the  partition function for closed string states with fixed size, mass,
   charge is
   \be \label{numberfixedNRcharge} 
    G_c    \sim {2\ov \Gamma({d \ov 2})}
     \left({3 \sqrt{d-1} \ov 8\pi^2 \mathcal{N}} \right)^{{d \ov 2}}
    \left({R \ov N_L^{{1 \ov 4}}N_R^{{1 \ov 4}}}\right)^{d-1} 
     {e^{\pi\sqrt{d-1}\left(\mathcal{N}-{3 \ov 8\pi^2\mathcal{N}}R^2\right)}\ov N_L^{{3 \ov 4}}N_R^{{3 \ov 4}}}
   \nonumber \\ 
   \ee
  \item and the entropy
   \be
     S & = & \ln(G_c) \nonumber \\
      & \sim & \pi \mathcal{N}\sqrt{d-1}-{3\sqrt{d-1} \ov 8\mathcal{N} \pi}R^2 
   +\ln\left({R^{d-1} \ov N_L^{{d+2 \ov 4}}N_R^{{d+2\ov 4}}\mathcal{N}^{{d \ov 2}}} \right)
   \ee
  \end{itemize}
 with
  \be
     R^2_{\text{max number}} & = &   {4\pi \,\sqrt{d-1}  \ov 3} \mathcal{N} \\
     R^2_\text{average} & = &  {4\pi \,d  \ov 3} {\mathcal{N} \ov \sqrt{d-1}}
  \ee
 \item for the {\bf bosonic string}
 \begin{itemize}
  \item the  partition function for closed string states with fixed size, mass,
   charge is
   \beq \label{numberfixedNRboscharge}
    G_c = {2 \ov \Gamma({d \ov 2})}
     \left({\sqrt{3 (d-1)} \ov 4\sqrt{2}\pi^2 \mathcal{N}} \right)^{{d \ov  2}} 
    \left({R \ov N_L^{{1 \ov 4}}N_R^{{1 \ov 4}}}\right)^{d-1} 
     {e^{\pi\sqrt{d-1}\left(\sqrt{{2 \ov 3}}\mathcal{N}-{\sqrt{3} \ov 4 \sqrt{2}\,\pi^2\, \mathcal{N}}  R^2\right)}\ov N_L^{{3 \ov 4}}N_R^{{3 \ov 4}}}
   \eeq
  \item and the entropy
   \be
     S & = & \ln(G_c) \nonumber \\
      & \sim & 2\pi \mathcal{N}\sqrt{{d-1\ov 6}}-\!\!
      {\sqrt{3\,(d-1)} \ov 4\sqrt{2}\,\mathcal{N}\, \pi}  R^2 + \!\!
   \ln\!\!\left({R^{d-1} \ov N_L^{{d+2 \ov 4}}N_R^{{d+2 \ov 4}}\mathcal{N}^{{d \ov 2}}}\!\!\right).
   \ee
  \end{itemize}
\end{itemize}

\subsubsection{BPS states} \label{BPSentropyfree}

We study, now, BPS configurations of fundamental superstrings.
They are states with:
 \beq
 M^2=Q^2_L, \quad N_L=0, \quad N_R=\sum_i n^im^i.
 \eeq
We find:
 \begin{itemize}
  \item the  partition function for closed BPS string states with
    fixed size, mass, charge 
   \be \label{numberfixedNRBPS}
    G_c = {2\ov \Gamma({d \ov 2})}
     \left({3 \sqrt{d-1} \ov 8\pi^2 \sqrt{N_R}} \right)^{{d \ov 2}}
    \left({R \ov N_R^{{1 \ov 4}}}\right)^{d-1} 
     {e^{\pi\sqrt{d-1}\left(\sqrt{N_R}-{3 \ov 8\pi^2\sqrt{N_R}}R^2\right)}\ov N_R^{{3 \ov 4}}}.
  \nonumber \\ 
   \ee
  \item and the entropy
   \be
     S & = & \ln(G_c) \nonumber \\
      & \sim & \pi \sqrt{N_R}\sqrt{d-1}-\!\!{3\sqrt{d-1} \ov 8\sqrt{N_R} \pi}R^2\!\! 
   +\ln\left({R^{d-1} \ov N_R^{{d+1 \ov 2}}}\!\!\right).
   \ee
  \end{itemize}

It is interesting to note that the average radius for this
ensemble is
 \beq \label{sizeBPS}
 R^2_{\text{average}} = {4\pi \,d  \ov 3} \sqrt{{N_R \ov d-1}}.
 \eeq
which is larger than the Schwarzshild radius at transition/matching point (see
section \ref{BPSstatescorrections}).

\section{Size distribution for highly excited string states with self-interactions} \label{massshiftsec}

The counting of states at a given mass level is affected by
the self-interaction of the string, unless we are considering  
supersymmetric configurations, which enjoy a protection
mechanism for the mass. 

The idea is that the formulas
for the entropy
obtained in sections \ref{sizesec} and \ref{stateswithcharges}
will receive corrections due to renormalization of the mass and size, 
such that the partition function will be dominated by
string states with a typical
size within the Schwarzschild radius of the correspondent
black hole at the 
transition/matching point.

The questions we want to address are:
 \begin{itemize}
  \item what is the distribution of string states at a certain mass
    and charge in terms of the size at non-zero coupling? Are small
    sizes preferred?
  \item how does the value of the Schwarzschild radius emerges from the
    string point of view\footnote{We know it is important because we
      are following {\em two} descriptions of the same system,
      according to the correspondence principle: the classical black
      hole and the quantum string ensemble.The Schwarzschild radius
    is characteristic of the classical description; does it arise as
    a special quantity also from the point of view
    of self-interacting strings? And how?}?
  \item what is the minimal size a string state can attain at 
      non-zero coupling?
 \end{itemize}

In order to answer those questions we need to compute the
effect of interactions on string states. These will modify
the dynamical equations of the states and their partition
function. We will study self-energy (mass-renormalization) corrections
for string states 
elaborating on \cite{chialvamassshift}.
For simplicity, we consider $|Q^i_L|=|Q^i_R|, \,\, N_L=N_R$ and define
$Q^i \equiv Q_L^i$.

Formally, in operatorial form, 
the average squared mass-shift for states constrained in both mass, charges and
size
would be obtained from the formula
 \begin{align} \label{masshisftfixedNR} 
 \overline{\Delta M^2}_{|_{N,Q,R}} & =  G_c(N,Q,R)^{-1}  \\ 
  & \times \!\!
  \text{tr}[\widehat{\Delta M}^2\,\delta(N-\hat N_R)\,\delta(N-\hat N_L)\delta(Q_R-\hat Q_R)\delta(Q_L-\hat Q_L)\delta(R-\hat R)]
 \nonumber
 \end{align}
where $\widehat{\Delta M}^2$ is an operator yielding the squared mass
shift once 
applied to a  
set of states\footnote{It could be obtained opportunely normalizing the
real part of the one-loop S-matrix operator.}. 

Integrating over $R$ and dividing by $G_c(N,Q)^{-1}$, 
formula (\ref{masshisftfixedNR}) translates into
 \beq \label{massshiftdistr}
  \overline{\Delta M^2}_{|_{N,Q}} = 
   \int d R \,\,\overline{\Delta M^2}_{|_{N,Q,R}} \,\,\rho_c(N, Q, R).
 \eeq
where $\rho_c(N, Q, R)$ is the {\em density} of string states at given
mass-level, charge and size, equal to
 \beq \label{densityNQR}
  \rho_c(N, Q, R) \equiv {G_c(N, Q, R) \ov G_c(N, Q)} \,.
 \eeq
and $\overline{\Delta M^2}_{|_{N,Q}}$ is the average squared mass-shift at
fixed $N, Q$\footnote{$\overline{\Delta M^2}_{|_{N,Q}}= G_c(N, Q)^{-1} \, 
  \text{tr}[\widehat{\Delta M}^2\, \delta(N-\hat N)\delta(Q_R-\hat
    Q_R)\delta(Q_L-\hat Q_L)]$, \,$Q^i=Q_L^i$.}.

In \cite{chialvamassshift} it was obtained\footnote{In
  \cite{chialvamassshift} $g_s$ was redefined to get rid of a 
  positive constant of order one in the result of the average squared
  mass shift. However, we can still retain the formula $G_N\sim g_s^2
  (\alpha')^{{d-1 \ov 2}}$, relating Newton's constant to the string
  coupling for small curvature radii  $r_i \sim \sqrt{\alpha'}$ of
       compactified dimensions, since at this level we do not pay
       attention to constant factors of order one. This implies that
       we will not be able to account for the specific proportionality
factor of ${1 \ov 4}$ in Beckenstein's formula for the entropy.}  
 \beq \label{masshisftconcl}
  \overline{\Delta M^2}_{|_{N,Q}} = -g_s^2 (M_0^2-Q^2)^{1+{3-D \ov 4}} \, ,
 \eeq 
where we have used the definition of the tree-level mass
 \beq
  M_0^2 = N+Q^2.
 \eeq
It is possible to see that the squared mass-shift becomes
non-negligible (of order one)
for
 \beq \label{couplimportself}
  g_{se} \sim (M_0^2-Q^2)^{{d-6 \ov 8}}, 
 \eeq
in analogy with the expectations form  the field theory argument in
formula 3.3 in \cite{HorPolSelf} (they consider only $Q^2=0$).

In order to solve equation (\ref{massshiftdistr}), it is useful to
propose an ansatz.
As it was discussed in section 5 of \cite{chialvamassshift}, the
formula (\ref{masshisftconcl}) is constituted by two factors with
different origins. In particular, the factor $(M_0^2-Q^2)^{{3-D \ov 4}}$
is related to the spatial range of the interaction which was found to
provide the dominant contribution to the mass-shift (namely
gravitational interactions). In  \cite{chialvamassshift}, 
it was also discussed how this was in fact
given by the average length of the massive string (we have indeed
found $R^2_\text{average} \sim \sqrt{M_0^2-Q^2}$ in sections
\ref{sizesecfinform}, \ref{nonBPSpartfunc}). 

We therefore 
consider the ansatz
 \beq \label{ansatz}
  \overline{\Delta M^2}_{|_{N,Q,R}} = -g_s^2 c (M_0^2-Q^2)\, R^\alpha , 
 \eeq
with $c$ a suitable proportionality constant.

From
(\ref{massshiftdistr}, \ref{densityNQR}, \ref{masshisftconcl})
and the results 
for $G_c(N, Q, R)$ obtained in sections \ref{sizesecfinform},
\ref{nonBPSpartfunc}, we determine the correct power $\alpha$
 \beq
  \alpha = 2-d \, ,
 \eeq
so that\footnote{In this formula we write the mass-shift in
  terms of the true mass, which is probably even a more accurate
  estimate. Note also that this result for the correction is valid in
  perturbation 
  theory only for sizes larger than a certain minimal value. 
  Alternatively, it is possible to find solutions to
  the equation (\ref{massshiftdistr}) that deviates from 
  (\ref{aversqaurmasshiftsizes})
  at small $R$ and are valid for all sizes, such as 
  \beq \label{avermassshiftregularized}
   \overline{\Delta M^2}_{|_{M,Q,R}} = -g_s^2c {M^2-Q^2 \ov R^{d-2}}
    {2^{{d-2 \ov 2}} \ov \Gamma({d \ov 2})}
   \left(1-{\Gamma({d \ov 2},  3\sqrt{d-1}\,R^2\, (16\pi \sqrt{N})^{-1}) \ov \Gamma({d \ov 2})}\right)\, , 
  \eeq
where $c$ is the same as in (\ref{aversqaurmasshiftsizes}) and
$\Gamma(a, x)$ is the incomplete gamma function.

We prefer (\ref{aversqaurmasshiftsizes}) because
it is conceivable that perturbations theory breaks down for very
small sizes. However, the results we find remain in general true even using
(\ref{avermassshiftregularized}).}
 \beq \label{aversqaurmasshiftsizes}
   \overline{\Delta M^2}_{|_{M,Q,R}} = -g_s^2c (M^2-Q^2) R^{2-d} \, ,
 \eeq
with $ 
 c= {\Gamma({d \ov 2}) \ov \pi}
   \text{{\small$\left({16 \ov 3\sqrt{d-1}}\right)^{{d-2 \ov 2}}$}}
 $. In the following we set $c=1$. There is no loss of generality in
   doing this because we can opportunely redefine $g_s$. Indeed, we
   are not going to pay attention to factors of order one.

Note that our result is different from that in \cite{DamVenSelf} in
the power of the mass (they consider only $Q^2=0$). 
As we said in section \ref{principle}, around
(\ref{damvenmassshift}), their result would imply 
that we cannot apply perturbation
theory uniformly on the whole string spectrum in lower dimensions,
certainly not in the 
limit of large masses. With the
result (\ref{masshisftconcl}), obtained from well-defined string
amplitudes in \cite{chialvamassshift}, 
perturbation theory is generally viable on the string
spectrum\footnote{This does not exclude that particular sets of
  states, not representing significant portion of the string spectrum
  (sets of ``measure zero'') and therefore not affecting the average,
  could have larger corrections and therefore not be suited for a
  perturbative treatment.}.

We can now discuss how the string distribution in terms of mass,
charge and size is modified by the corrections. The partition function
modifications proceed from\footnote{We consider adiabatic variations
  of the coupling, for which the entropy is unchanged, so that
$\log(G_c(M_0, Q, R)) = \log(G_c(M^2, Q, R))$.} 
 \beq
  G_c(M_0, Q, R_0) = G_c(M^2, Q, R)\, ,
 \eeq
where we neglect renormalization of $R$ and use
 \beq
  M_0^2=M^2-\overline{\Delta M^2}_{|_M,Q,R}
 \eeq
from the definition of mass-shift. This translates
into\footnote{The winding number and also the Kaluza-Klein mode (T-dual
  to it) are not renormalized.}
 \beq
  M_0^2-Q^2= M^2-Q^2+g_s^2 (M^2-Q^2) R^{2-d}.
 \eeq
The important effects in $G_c(M^2, Q, R)$ arise from the exponential
factor (see (\ref{numberfixedNRcharge})), which now at leading
order\footnote{We neglect 
renormalization of $R_0$, that is we reckon that 
$\overline{\Delta R^2}_{|_{M,Q,R}} \ll \overline{\Delta M^2}_{|_{M,Q,R}}$.} is
 \beq
  e^{2\pi\sqrt{d-1}\left(\sqrt{M^2-Q^2}+g_s^2 {\sqrt{M^2-Q^2}\ov 2 R^{d-2}}-{3 \ov 32\pi^2\sqrt{M^2-Q^2}}R^2\right)}
 \eeq
We can see from this that the behavior of the string partition
function and
the entropy (its logarithm)
changes for 
 \beq 
  g_s^2 {\sqrt{M^2-Q^2}\ov 2 R^{d-2}} \geq 1 \,
 \eeq
when it becomes dominated by strings of size $R \lesssim R_b$, where
 \beq
  R_b \sim (g_s^2 \sqrt{M^2-Q^2})^{{1 \ov d-2}} \,.
 \eeq
But $R_b$ is indeed the value of the Schwarzschild radius
for a charged black hole in $d$ spatial
dimensions\footnote{\label{gravdress} We
  consider here black holes obtained by the usual procedure of lifting
  a $D=d+1$ dimensional Schwarzschild solution to $D+1$ dimensions,
  boosting along the new extra dimension and reducing down to
  $D$ dimensions again (see \cite{Peet:2000hn}). $R_b$ is the value of the
  horizon radius $r_E$ in the Einstein frame
  when $M$ is identified with the ADM mass
  $M_{BH}$ (see \cite{HorPolSelf}, for $d=3$, their result can be
  extended to $d > 3$. Consider that we have $|Q^i_L|=|Q^i_R|$). We
  define the horizon radius in the Einstein frame as 
   $r^2_E = e^{-{4 \ov D-2}\phi}r^2_S $ where $\phi$ is the dilaton and $r_S$ is
  the horizon radius in the string frame. In this way the area of the black
  hole horizon is $A \sim r_E^{D-2}$. Of
  course, in view of the correspondence principle, the 
  identification $M_{BH} = M$ is done at a specific
  value of the coupling. In particular, the matching in this
  case should not be done when $r_E\sim l_s$, but when $r_S \sim l_s$
  (see \cite{HorPolSelf}).}. In the 
description of our string distribution, it becomes important when
$R_b \geq l_s$, where $l_s$ is the string length, which occurs at $g_s
\sim (M^2-Q^2)^{-{1 \ov 4}}$, in accordance with the correspondence
principle\footnote{This is strictly true when $Q^2=0$: as we said in
  footnote \ref{gravdress}, when charges are present, the match occurs
  at $r_S \sim l_s$. However, if we trust the
  classical description further up when $r_E \sim l_s$, which is
  sensible when $N=M_o^2-Q^2 \gg 1$, then
  $G_N \sim \sqrt{M^2-Q^2}$ there.} in \cite{HorPolSelf}. 
The value $R^2 \sim \sqrt{M^2 -Q^2}$ remains, in any case,
at small $g_s$, a local maximum of the partition function/entropy.

Another interesting question that we can try to ask ourselves at this
point regards the minimal value for the size of a string state with
self-interaction. 
We can estimate it from the modification to the equation of motion:
for a state $|\phi\rangle$ of true squared mass $M^2$
the energy is
 \beq \label{Kleinordonselfinteract}
  E_\phi^2=\vec p^2+M^2=\vec p^2+N -g_s^2 {M^2-Q^2 \ov R^{d-2}}\, .
 \eeq

We
can make a rough estimate for the minimal radius 
(and this suffices for our work here),
as usual in quantum mechanics, through the Heisenberg
uncertainty principle. That is, we roughly expect\footnote{In our
  units $\hbar=1$.} 
$p^i \sim {1 \ov R}$ and then
we estimate the energy of the minimum $E_0$ by minimizing
(\ref{Kleinordonselfinteract}) with respect
to $R$.
 
We immediately see the difference between cases $d=3$, $d=4$,
$d \geq 5$. For $d=3$ we find a lowest state for 
 \beq
  R = {2 \ov g_s^2 (M^2-Q^2)}
 \eeq
for $d \geq 5$ instead there is no lowest state, and apparently the
energy has no 
lower bound. This would signal an instability. But it actually occurs
outside the domain of validity of perturbations theory, since it would
show up when $N -g_s^2 {M_0^2-Q^2 \ov R^{d-2}} < 0$ and therefore when
$\overline{\Delta M^2}_{|_M,Q,R} > M_0^2$. This means that in this
case we cannot
trust our first order perturbative corrections to give us an
exhaustive insight on 
what happens for extremely compact string states.
For $d=4$ the energy is negative past a critical coupling.

\subsection{Matching point for BPS states}\label{BPSstatescorrections}

The horizon radius (and therefore the entropy) for the BPS black
holes corresponding to the states
considered here vanishes. The string entropy, instead, as shown in  
section \ref{BPSentropyfree}, is non-zero and grows with 
the mass. It would seem that the string/black hole correspondence
principle does not work in this case. 

Sen's proposal \cite{Sen:1995in}
at this point was that the entropy formula to be considered should
involve the area of the surface at the ``stretched horizon'' and not
the Schwarzschild one. The rationale behind this is that for BPS
black holes the curvature of the classical geometry is comparable to
the string scale already at the stretched horizon and
therefore the classical description is unreliable already there,
where string effects show up. Following this line of reasoning, in a
class of solutions it was shown that a higher derivative corrected
computation of entropy matches the string entropy \cite{Dabholkar:2004yr}.

We know that the counting of BPS states, 
and therefore their
entropy,  for fixed mass and
charge do not receive corrections due to the vanishing of their
two-points torus amplitude. Nevertheless, it would be interesting to study
the configuration of the string, since the matching is at non-zero
string coupling.

What is the nature of the corrections that would favor more
compact states is not obvious. 
In \cite{Cornalba} it was conjectured a correction
whose form was very similar to the one obtained from one-loop
self-energy corrections; however, it was not clarified which kind of
string diagram would produce it.

\section{Discussion and Conclusions}

The main new results of this paper are twofold: 1) the computation of the
partition function for closed very massive
free string states in the microcanonical
ensemble at given (large tree-level) mass, charges (Neveu-Schwarz) and size,
2) the study of the dynamics of the string in presence of
self-interactions and the resulting modifications of the partition
function at given (large true) mass, charges and size.

At tree-level, our quantum computation shows 
how the (expected) random walk picture for the
string arises. On the other hand, the
self-interaction of the string modifies the distribution of the
string states in an important way\footnote{We will not discuss here
  the details involved in computing and 
analyzing (\ref{masshisftconcl}), but we direct the reader to
\cite{chialvamassshift}. }. 
These results allows us to clarify the correspondence 
(\cite{corresp, HorPol96}) between 
string states an black holes in non-supersymmetric configurations.
 
In particular, our results 
make many of the conclusions in \cite{HorPolSelf}
(we think) physically
clearer by directly computing the relevant quantities for
a well-defined string ensemble. The computations, indeed, are
performed rigorously in string theory formalism in the asymptotic
limit of large string masses 
(highly excited states). In comparing our results with those in
\cite{HorPolSelf}, note however that we perform our analysis using a
microcanonical ensemble of closed strings
and not a canonical one for open strings as in \cite{HorPolSelf}.

When our results differ from  \cite{HorPolSelf}, they
make even more compelling the existence of a correspondence
(complementarity) between strings and black holes. For example, 
 \cite{HorPolSelf} seems to find an instability for $d=5$ and, from the string
 $\to$ black hole side of the correspondence, the string seems to
 collapse to a black hole at a value for the coupling lower than the
 critical one expected from the correspondence principle 
(which was found investigating the  
black hole $\to$ string side of the correspondence). Also,
  \cite{HorPolSelf} suggested that  in $d > 6$ most excited string states
 would never correspond to a black hole, at any value of the coupling.

The physical picture that emerges form our computations, shows instead
that at sufficiently large 
coupling $g_s$, in any dimensions the string ensemble will be
dominated by typical strings of size\footnote{To simplify the
  notation here we take $Q^2=0$.} $R \lesssim (g_s^2 M)^{{1 \ov d-2}}$
which corresponds to the black hole Schwarzschild
radius and which is of the order of the
string scale precisely at the expected correspondence point $g_s \sim
M^{-{1 \ov 2}}$. We find that  for $d > 6$, this occurs when the average
self-gravity correction is not yet strong (see
(\ref{masshisftconcl}, \ref{couplimportself})). 
The details are different between dimensions $d=3$, $d=4$ and
$d \geq 5$. 

Let us focus on the best understood
case ($d=3$). The entropy of the self-gravitating string 
is dominated by the lowest bound state and is given by
 \beq
  S(M, Q) \sim 2\pi\sqrt{d-1}\sqrt{M^2-Q^2}
      \left(1+{g_s^4 (M^2-Q^2)\ov 4} \right).
 \eeq
Let us consider now for simplicity the case $Q^2=0$ and 
obtain the temperature of ensemble of closed string by
differentiating with respect to $M$. We find:
 \beq
  T \sim T_H\left(1 - {3 \ov 4} g_s^4 M^2 \right), \qquad 
   T_H ={1 \ov 2\pi\sqrt{d-1}}
 \eeq
which shows the diminishing of the Hagedorn
temperature\footnote{Reinstating $\alpha'$, indeed, 
$T_H ={1 \ov \pi\sqrt{d-1}\sqrt{\alpha'}}$\,, which is the Hagedorn
  temperature of type II superstring.} $T_H$ in
agreement with \cite{HorPolSelf}. Note however that we are not
describing our system in a canonical ensemble and therefore the
discussion of phase transitions is different from the one in canonical
formalism. We will not deal with these interesting questions here, but
leave them for future investigation.

\section{Acknowledgments}

I would like to thank Ulf Danielsson and Paolo Di Vecchia for the
many conversations about this project and Tibault Damour for the
discussions 
regarding the problems
treated in it. I am especially grateful to Roberto Iengo for his
useful remarks in 
discussing the results of this work.



\end{document}